\documentclass[aps,prl,reprint,superscriptaddress]{revtex4-1}
\usepackage{amsmath}
\usepackage[euler]{textgreek}
\usepackage{graphicx}
\begin{document}

% Use the \preprint command to place your local institutional report
% number in the upper righthand corner of the title page in preprint mode.
% Multiple \preprint commands are allowed.
% Use the 'preprintnumbers' class option to override journal defaults
% to display numbers if necessary
%\preprint{}

%Title of paper
\title{Density modulation-induced absolute laser-plasma-instabilities: simulations and theory}

% repeat the \author .. \affiliation  etc. as needed
% \email, \thanks, \homepage, \altaffiliation all apply to the current
% author. Explanatory text should go in the []'s, actual e-mail
% address or url should go in the {}'s for \email and \homepage.
% Please use the appropriate macro foreach each type of information

% \affiliation command applies to all authors since the last
% \affiliation command. The \affiliation command should follow the
% other information
% \affiliation can be followed by \email, \homepage, \thanks as well.
\author{J. Li}
%\email[]{Your e-mail address}
%\homepage[]{Your web page}
%\thanks{}
%\altaffiliation{}
\affiliation{Department of Mechanical Engineering and Laboratory for Laser Energetics, University of Rochester, Rochester, New York 14627, USA}

\author{R. Yan}
%\email[]{Your e-mail address}
%\homepage[]{Your web page}
%\thanks{}
%\altaffiliation{}
\affiliation{Department of Mechanical Engineering and Laboratory for Laser Energetics, University of Rochester, Rochester, New York 14627, USA}

\author{C. Ren}
\email{Corresponding author.\\chuang.ren@rochester.edu}
%\homepage[]{Your web page}
%\thanks{}
%\altaffiliation{}
%\thanks{A footnote to the article title}
\affiliation{Department of Mechanical Engineering and Laboratory for Laser Energetics, University of Rochester, Rochester, New York 14627, USA}
\affiliation{Department of Physics and Astronomy, University of Rochester, Rochster, New York 14627, USA}

%Collaboration name if desired (requires use of superscriptaddress
%option in \documentclass). \noaffiliation is required (may also be
%used with the \author command).
%\collaboration can be followed by \email, \homepage, \thanks as well.
%\collaboration{}
%\noaffiliation

\date{\today}

\begin{abstract}
% insert abstract here
Fluid simulations show that when a sinusoidal density modulation is superimposed on a linear density profile, convective instabilities can become absolutely unstable. This conversion can occur for  two-plasmon-decay and stimulated Raman Scattering instabilities under realistic direct-drive inertial confinement fusion conditions and can affect hot electron generation and laser energy deposition. Analysis of the three-wave model shows that  a sufficiently large change of the density gradient in a linear density profile can turn convective instabilities into absolute ones. An analytical expression is given for the threshold of the gradient change, which depends on the convective gain only. 
\end{abstract}

% insert suggested PACS numbers in braces on next line
\pacs{}
% insert suggested keywords - APS authors don't need to do this
%\keywords{}

%\maketitle must follow title, authors, abstract, \pacs, and \keywords
\maketitle

% body of paper here - Use proper section commands
% References should be done using the \cite, \ref, and \label commands
%\section{}

 Success of inertial confinement fusion (ICF) requires a comprehensive understanding of laser-plasma instabilities (LPI). Linear theory is essential to assess the importance of various LPI in ICF.  Absolutely unstable modes are deemed important since they can grow to large amplitudes before nonlinear effects set in. For convectively unstable modes, their amplitudes are capped by the Rosenbluth convective gain formula \cite{Rosenbluth1972} . Linear theory is usually developed assuming smooth plasma density profiles and the resultant gain formulas are used in LPI-assessing codes \cite{Strozzi2008}.  However, during a ns-long laser pulse, significant density modulations from various LPI modes can develop and affect the subsequent laser-plasma interactions.  LPI under density modulations can be significantly different from those under smooth density profiles. Previous study found that density modulations can reduce the LPI growth rates in homogeneous plasmas but can change convective instabilities into absolute ones in inhomogeneous plasmas \cite{Nicholson1976,Nicholson1974,Laval1976}. Pickard and Johnston found that when the density profile deviates from the linear profile a strong enough laser can drive absolute modes but for short-scale-length targets the required laser intensity is too high to be a concern \cite{Picard1983}. Up until now, the density modulations-induced absolute modes had not been observed in experiments or simulations under realistic ICF conditions  and no theory existed to predict the growth rates when this conversion occurs.

In this Letter, using fluid simulations, we show for the first time two instances that under realistic ICF conditions and density modulations this convective-to-absolute mode conversion is important. One is the two-plasmon decay instability (TPD) \cite{Rosenbluth1972,Rosenbluth1973,Liu1976} in conventional  direct-drive ICF. These TPD modes have low phase velocities and are essential to the staged-acceleration mechanism for hot electron generation \cite{Yan2012}.  The other is the Stimulated Raman Scattering (SRS) instability \cite{Drake1974,Forslund1975} in the ignition stage of shock ignition \cite{Betti2007}. These SRS modes are important to generating hot electrons with moderate energy that can assist shock generation in shock ignition \cite{Theobald2012, Betti2007} and can explain the intermittent LPI activities in the simulations \cite{Yan2014}. 

Using a two-straight-line density profile,  we further show that this conversion is due to a change of the density gradient and derive an analytical formula for the gradient change threshold that depends on the convective gain only.  This formula can also apply to the more general case where a sinusoidal density modulation is superimposed on a linear density profile. It shows that due to the long scale length in current ICF experiments, the density modulation-induced absolute modes are an important part of LPI. Growth rates of these absolute modes can also be calculated from this model.

We now show that a sinusoidal ion density modulation can turn convective TPD modes into absolute ones under realistic ICF conditions. For a linear density profile, the theory \cite{Liu1976,Simon1983} predicts TPD are absolutely unstable only in a narrow density region close to the $1/4 n_c$ surface and only convectively unstable below this absolute zone. For direct drive ICF, TPD-generated hot electrons can cause target preheat \cite{Smalyuk2008, Goncharov2008}. On the other hand, in the ignition phase of shock ignition (SI) \cite{Betti2007}, hot electrons with moderate energy (less than $\sim100$ keV) could be stopped at the compressed shell and help drive the ignition shock. A recent study\cite{Yan2012} found that hot electrons are stage-accelerated first by the TPD modes in the $n=0.23n_c$ ($n_c$ is the critical density) region where the theory predicts only convective TPD modes, and then by those in the higher density region ($n=0.245n_c$) where the theory predicts absolute modes. The modes in the convective zone have lower phase velocities and can more efficiently accelerate thermal electrons. Significant TPD modes in the convective zone were also observed to be energetically dominant in previous OMEGA experiments \cite{Seka2009} and PIC simulations \cite{Yan2009}. These modes were found to correlate to ion density fluctuation \cite{Yan2012}. However, why these modes can grow beyond the convective gain has not been understood. 

\begin{figure}
\includegraphics[width=0.5\textwidth]{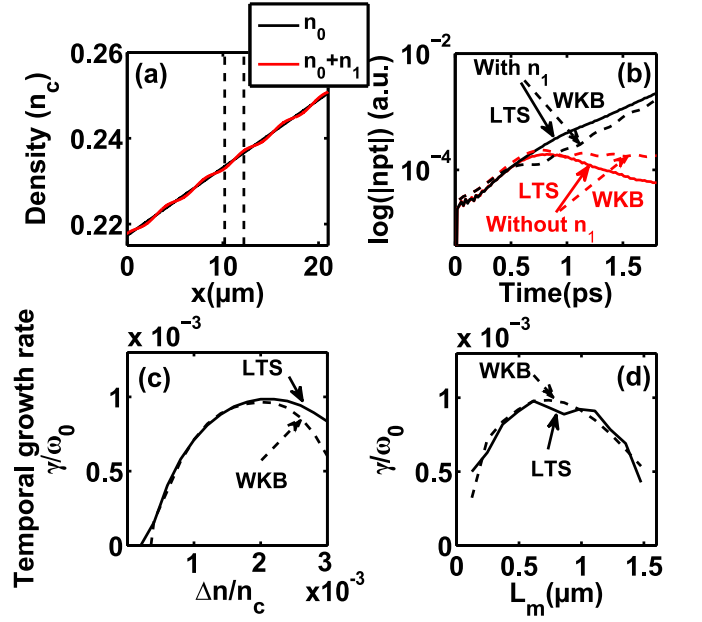}% Here is how to import EPS art

\caption{\label{fig:fig1} (color online). (a) The density profiles with and 
without sinusoidal density modulation; (b) The time evolution of the TPD modes
 in LTS and WKB simulations; TPD temporal growth rates of the absolute
 instability (c) with different $\Delta n$ for $L_m=0.65 \mu m$ 
and (d) with different $L_m$ for $\Delta n=0.002n_c$ 
 }
\end{figure}

We use the LTS code \cite{Yan2010}, which solves the full set of linear TPD equations, 
\begin{eqnarray}
\frac{\partial \Psi}{\partial t}=\phi-3v_e^2\frac{n_p}{n_0}-\roarrow v_0\cdot\nabla\Psi\nonumber\\ 
\frac{\partial n_p}{\partial t}=-\nabla\cdot(n_0\nabla\Psi)-\roarrow v_0\cdot\nabla n_p,\\
\nabla^2\phi=n_p.\nonumber
\end{eqnarray}
Here $ \Psi, \phi $ and $n_p$ are the linearized electron velocity potential, electric field potential and electron density perturbation, respectively. And $n_0$ and $\roarrow v_0$ are the background plasma density and electron oscillation velocity in the laser field. LTS has been successfully benchmarked with the TPD theory \cite{Simon1983, Afeyan1997} in the linear density profile case \cite{Yan2010} but it can also study TPD growth under an arbitrary $n_0$.  We present two LTS simulations, one with and the other without an ion density modulation, to illustrate the transition from convective TPD modes to absolute modes. The physical parameters in the LTS simulations are chosen for typical OMEGA experiments. Without the ion density modulation, the background ion density profile $n_0$ is a smooth linear function, as shown by the black solid curve in Fig.\ref{fig:fig1}(a). It ranges from 0.217 $n_c$ to 0.253 $n_c$ with a scale length $L= 150 \mu m$ at the $1/4 n_c$ surface. The electron temperature is $T_e = 3$ keV. The laser pulse is a plane wave coming from the left side in Fig. 1(a) with an intensity $ I = 6 \times 10^{14} W/cm^2 $ and a wavelength of 0.33 $\mu m$.  With the ion density modulation, the background ion density profile is the sum of $n_0$ and an oscillation component $n_1=\Delta n\sin(x/L_m)$, as shown in Fig. 1(a) by the red line. The modulation amplitude $\Delta n$  and characteristic length $L_m$ are chosen from typical values in the nonlinear steady state of the PIC simulations with the same parameters\cite{Yan2012}. Here we choose $\Delta n=6\times10^{-4} n_c$ and $L_m=0.65\mu m$. For both simulations,  the box size is $400\times 819.2 c/\omega_0$ $(21.2 \times 43.5 \mu m)$ with a grid of $2000 \times 4096$ and a grid size $\Delta x = 0.2 c/\omega_0$ $(0.01 \mu m)$, where $c$ is the speed of light and $\omega_0$ the laser frequency. The time step is $\Delta t = 0.1414 \omega_0^{-1}$ $(0.025 fs)$.  The boundary condition is periodic in the transverse direction, and dQ/dx = 0 in the longitudinal direction, where Q stands for $ \Psi, \phi $ and $n_p$. The initial electron density perturbation seed is set as a 2D Gaussian function with a full width at half maximum (FWHM) of $0.12 \mu m$ in each direction and an amplitude of $10^{-6}$ in the middle of the simulation box at $n_0= 0.235 n_c$. We focus on the TPD mode with a transverse wave number $k_y= 0.48 \omega_0/c$ inside a narrow region with a width of $2 \mu m$ and centered at $0.235n_c$ (between the two black dashed lines in Fig. 1(a)). Theory for TPD in homogeneous plasmas \cite{Kruer} predicts this is the resonant mode for this density. Due to the density gradient this mode is convective under the linear $n_0$.  

Figure 1(b) shows the time evolution of the plasma wave amplitude of the $k_y= 0.48\omega_0/c$ mode in this region with and without the ion density modulation $n_1$. Without $n_1$, the plasma wave first grows and then saturates after 0.8 ps, consistent with the theory prediction \cite{Rosenbluth1973}. By contrast, with $n_1$, the plasma wave continues to grow exponentially and turned from convective to absolute. 

To isolate the essential physics and develop a theory, we further simplify Eqs.(1) into the standard 3-wave model under the WKB approximation \cite{Yan2010},
\begin{eqnarray}
(\frac{\partial }{\partial t}+V_1\frac{\partial }{\partial x})a_1=\gamma_0a_2e^{i\phi(x)},\nonumber\\ 
(\frac{\partial }{\partial t}+V_2\frac{\partial }{\partial x})a_2=\gamma_0a_1e^{-i\phi(x)},\\
\nonumber
\end{eqnarray}
where $\gamma_0=-k_y^2v_0^2(\omega_2/\omega_1-k_2^2/k_1^2)(\omega_1/\omega_2-k_1^2/k_2^2)/16$ \cite{Yan2010} is the TPD temporal growth rate in a homogeneous plasma, $a_{1,2}$, $V_{1,2}$, $\omega_{1,2}$ and $k_{1,2}$ stand for the amplitudes, group velocities, frequencies and wave numbers of the two plasmons. The phase mismatch $\phi(x)=\frac{1}{2}{d\kappa(x)}/{dx}=\frac{1}{2}\kappa'x^2$ is for a smooth linear density profile, where $\kappa(x)=k_0(x)-k_1(x)-k_2(x)$ is the wave vector mismatch. For the same parameters in the LTS simulations, $V_1=0.0375 c, V_2=-0.0076 c, \gamma_0=0.00147\omega_0$, and $\kappa'=0.0093(\omega_0/c)^2$. With the same sinusoidal ion density modulation $n_1$ in the LTS simulations, $\phi(x)=\frac{1}{2}\kappa'x^2+k_mL_m[1-cos(x/L_m)]$, where $k_m=0.052\omega_0/c$ is the wave vector mismatch caused by $\Delta n=6\times10^{-4} n_c$. We numerically solve Eqs.(2), which is referred as the WKB simulations hereafter, with the same seeding in the LTS simulation, and a spatial grid of $0.1 c/\omega_0$ , a time step of $0.0999/\omega_0$, and a simulation box size of $5000 c/\omega_0$, which is large enough so that $a_1$ and $a_2$ do not reach the boundary in the whole simulation. 

The WKB and LTS simulation results reasonably agree [Fig. 1(b)]. Using the LTS and WKB simulations, we scan a wide range of $\Delta n$ and $L_m$ that typically occurs in the PIC simulations, and plot the measured absolute growth rates in Fig 1(c)(d), which again show that the WKB model retains the essential physics.  The absolute growth rate maxes at $\Delta n \approx 2\times10^{-3}n_c$ and $L_m \approx 0.65\mu m$, and the maximum growth rate is $0.7 \gamma_0$. This is more than three times of the conventional absolute modes growth rates near the $1/4 n_c$ region\cite{Simon1983}. They are more effective in generating hot electrons due to their lower phase velocities. In addition, since they locate in the lower density region, they can also affect the LPI close to the $1/4n_c$ surface through pump depletion. 

We have developed a theory for the conversion threshold and growth rate with a density gradient change model. We assume $V_1 > 0$ and $V_2 < 0$, which is the case for TPD and back scattering SRS, in  Eqs.(2). For a smooth linear density profile $n_0(x)$, only convective modes exist\cite{Rosenbluth1973, Short2004} with a convective gain $exp{(\pi\Lambda)}$, where $\Lambda={\gamma_0^2}/{|\kappa'V_1V_2|}$.
  
\begin{figure}
\includegraphics[width=0.5\textwidth]{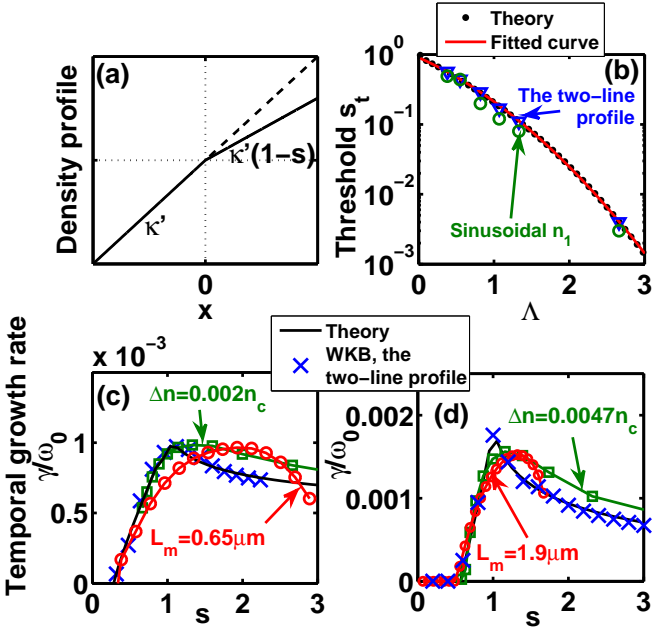}% Here is how to import EPS art

\caption{\label{} (color online). (a) The two-line density profile with reduced density gradient at $x>0$; (b) The comparison of the thresholds from Eq. (5), the fitted curve, and the WKB simulations with the two-line and sinusoidal profiles; The comparison of the absolute growth rates of the TPD (c) and SRS (d) cases from theory and the WKB simulations with the two-line and sinusoidal profiles}
%\caption{\label{fig:wide}Use the figure* environment to get a wide
%figure that spans the page in \texttt{twocolumn} formatting.}
%\texttt{twocolumn} formatting.}
\end{figure}

We consider a two-line density profile shown in Figure 2(a), leading to $\phi(x)=\frac{1}{2}\kappa'(1-s)x^2$, where $s=0$ for $x < 0$ and $s>0$ for $x>0$. The density gradient changes from $\kappa'$ to $\kappa'(1-s)$ for $x>0$. For this profile, we are seeking for a spatially bounded solution growing exponentially in time, which does not exist for unperturbed linear density profiles. Eqs.(2) can be analytically solved as an initial value problem with the initial condition of $a_1(x,t=0)=0, a_2(x,t=0)=a_{20}\delta(x)$. For $x<0$,  taking a Laplace transform in time and eliminating $a_2$, we can obtain \cite{Rosenbluth1973, ChambersPhD, Short2004}
\begin{equation}
\frac{\partial^2}{\partial z_-^2}a+(\frac{1}{2}-i\Lambda-\frac{1}{4}z_-^2)a=\frac{\gamma_0a_{20}\delta(x)}{V_1V_2}
\end{equation}
where $z_-=[\sqrt{\kappa'}x-{i}p({1}/{V_1}-{1}/{V_2})/{\sqrt{\kappa'}}]exp({i{\pi}/{4}}), a=a_1exp[-i\kappa'x^2/4+xp({1}/{V_1}+{1}/{V_2})/2]$ and $p$ is the Laplace transform variable . The solution well behaving at $-\infty$ is $\phi_-(x,p)=D_{i\Lambda-1}(iz_-)$, where $D$ is the parabolic cylinder function. For $x>0$, when $0<s<1$, the well behaved solution in this region is $\phi_+(x,s,p)=D_{-i\Lambda/(1-s)}(z_+)$ where $z_+=[\sqrt{\kappa'(1-s)}x-{i}p({1}/{V_1}-{1}/{V_2})/{\sqrt{\kappa'(1-s)}}]exp({i{\pi}/{4}})$ . Connecting the solutions at $x=0$, we obtain 
\begin{equation}
a(x,s,p)\propto \frac{\phi_-(x,p)\phi_+(0,s,p)\theta(-x)+\phi_+(x,s,p)\phi_-(0,p)\theta(x)}{W(s,p)}
\end{equation}
where the denominator $W(s,p)=\phi_+(0,s,p)\frac{\partial}{\partial x}\phi_-(0,p)-\phi_-(0,p)\frac{\partial}{\partial x}\phi_+(0,s,p)$. For a certain $s$, if $W(s,p)=0$ has a root $p_s$ with a positive real part $Re(p_s)>0$, $a(x,t)$ would grow exponentially with a growth rate of $Re(p_s)$. 

When $s>1$, the well behaved solution of $a(x,p)$ for $x>0$ is $\phi_{1+}(x,s,p)=D_{i\Lambda/(s-1)}(z_{1+})$ where $z_{1+}=[\sqrt{\kappa'(s-1)}x+{i}p({1}/{V_1}-{1}/{V_2})/{\sqrt{\kappa'(s-1)}}]exp({-i{\pi}/{4}})$. The corresponding form of the denominator $W_1(s,p)=\phi_{1+}(0,s,p)\frac{\partial}{\partial x}\phi_-(0,p)-\phi_-(0,p)\frac{\partial}{\partial x}\phi_{1+}(0,s,p)$. The equations $W(s,p)=0$ and $W_1(s,p)=0$ can be solved numerically to obtain the temporal growth rates. When $s=0$, $W(0,p)=\sqrt{\kappa'}exp({-i{\pi}/{4}-\pi\Lambda/2})$ is a constant and independent of p. No root $p_s$ can be found.

The density gradient change $s$ must be above a threshold $s_t$ for a positive growth rate to emerge. At the threshold, $Re(p_s)=0$. The imaginary part, $Im(p_s)$, physically represents the frequency of $a_1$. We can estimate if from the asymptotic solution of  $a_1\propto exp[i{\kappa'}{\alpha^2}(x-V_1t)^2/{2(1+\alpha)^2}]$, where $\alpha=-V_2/V_1$ \cite{Rosenbluth1973, ChambersPhD, Short2004}. The characteristic frequency is $\omega_{a1}=-Im(p_s)=\sqrt{{\kappa'}/{2}}V_1{\alpha}/{(1+\alpha)}$. Then the threshold condition is
\begin{equation}
\sqrt{1-s_t}\frac{D_{-i\Lambda/(1-s_t)}'(\frac{e^{i5\pi/4}}{\sqrt{2(1-s_t)}})}{D_{-i\Lambda/(1-s_t)}(\frac{e^{i5\pi/4}}{\sqrt{2(1-s_t)}})}=i\frac{D_{i\Lambda-1}'(\frac{e^{i7\pi/4}}{\sqrt{2}})}{D_{i\Lambda-1}(\frac{e^{7\pi/4}}{\sqrt{2}})}
\end{equation}
where the prime stands for the derivative of the parabolic cylinder function and the threshold $s_t$ only depends on $\Lambda$. The relation of $s_t$ and $\Lambda$ is plotted in Figure 2(b) and can be fitted with $s_t(\Lambda)=exp(-0.35\Lambda^2-1.1\Lambda-0.1)$. This shows even for a low $\Lambda=1$, $s_t=0.21$ is enough to induce absolute modes.

Numerically solving $W(p)=0$ ($0<s<1$) and $W_1(p)=0$ ($s>1$), we obtain the TPD growth rates vs $s$  (Fig. 2(c)). The results agree well with the direct numerical integration of Eqs.(2), validating the analysis. The growth rate maxes at $s=1$, corresponding to a homogeneous plasma at $x>0$ in Figure 2(a). The max growth rate at $s\approx1.0$ is close to the temporal growth rate of the absolute mode in a fully homogeneous plasma with the initial condition of a delta function \cite{Short2004}. When $s>1$, the absolute growth rate decays slowly.

For the case of density gradient increasing, i.e., $s<0$,  the model is equivalent to a gradient reduction of $|s|/(1+|s|)$ and a new gain parameter $\Lambda/(1+|s|)$. This shows that increasing the gradient is less effective to induce the absolute modes. For our TPD case,  the absolute instability generated by gradient increasing has a much higher threshold $s_t=-1.5$ and lower growth rate $\gamma=0.16\gamma_0$ for $s=-10$ compared with density gradient reduction where $s_t=0.29$ and $\gamma=0.7\gamma_0$ for $s=1$.

For the sinusoidal density modulation cases in Fig.1(a), we have not been able to develop a theory but the physics is similar. The modulations change the density gradient as well.  For the sinusoidal modulations, every combination of $\Delta n$ and $L_m$ corresponds to a gradient change by $s={k_m}/({L_m\kappa'})={\Delta n}/[{6v_e^2L_m\kappa'}({1}/{k_1}+{1}/{k_2})]$ where $v_e$ is the plasma thermal velocity. This allows us to compare the sinusoidal case with the two-line case. The threshold for the sinusoidal case in general depends on $k_m$ and $L_m$, not only on $k_m/L_m$, for a particular $\Lambda$. However, for each $\Lambda$, there exists a minimum modulation amplitude $k_m (\Delta n)$ for the conversion threshold. The combination of this $k_m$ and its corresponding $L_m$ gives an $s_t$ very close to the two-line model [Fig. 2(b)].  We also compare the absolute growth rates in Fig. 2(c). When $\Delta n$ is fixed and $L_m$ is allowed to vary (the case in Fig. 1(d)), the sinusoidal results agrees well with the two-line model when $s<1$ and has a 15$\%$ higher growth rate when $s>1$. When $L_m$ is fixed and $\Delta n$ is allowed to vary (the case in Fig. 1(c)) the agreement is less well but the maximum growth rate still agrees. Therefore the two-line model can be used to assess the maximum growth rates of the density-modulation-induced absolute modes for a given density profile. 

For SRS\cite{Kruer}, $\gamma_0=k_1v_0\sqrt{\omega_p^2/(\omega_1\omega_2)}/4$ in Eqs.(2).  Since one daughter wave is an electromagnetic wave with a much higher group velocity, the gain parameter $\Lambda$  is much smaller than in TPD for same laser and plasma conditions. For the particular case discussed above, SRS would not turn absolute under realistic density modulations. However, in the ignition phase of shock ignition, a new high-gain ignition scheme \cite{Betti2007}, the laser intensity is significantly higher and absolute SRS can be induced. At $n_e=0.22n_c$, the SRS match condition gives $V_1=0.0387c, V_2=-0.39c$ and $\kappa' =0.0021(\omega_0/c)^2$.  For a laser intensity of $I=2\times10^{15} W/cm^2 , \gamma_0=0.0033\omega_0$ and $\Lambda=0.38$. The growth rate vs $s$ for SRS in the two-line model is plotted in Fig. 2(d), showing a threshold at $s_t=0.56$ and a maximum growth rate of $1.76\times10^{-3}\omega_0$ at $s=1$. Two cases with the sinusoidal modulations, one with a fixed $L_m=1.9\mu m$ and the other with a fixed $\Delta n=4.7\times 10^{-3}n_c$ (typical values from the PIC simulation\cite{Yan2014}), are also plotted. They both show the existence of absolute SRS with peak growth rates of $1.5\times10^{-3}\omega_0$ . At shock ignition intensities, these density-modulation-induced absolute SRS can dominate over TPD at low densities since the later suffers stronger Landau damping. For example, for a $I=8\times10^{15} W/cm^2$ at $0.18n_c$,  $\Lambda=1$ for SRS. Our theory predicts a low threshold $s_t=0.21$ and a max growth rate of $0.0038\omega_0$. These SRS can suppress TPD near $n=0.25n_c$ through pump depletion. This qualitatively agrees with the experimental results \cite{Theobald2012}, where the backscattered SRS level increases significantly when the laser intensity increases. Moreover, these SRS modes' growth relies on the density modulations, which are first generated by the TPD and high frequency hybrid modes \cite{Afeyan1997} near $n=0.25n_c$. This coupling between the modes at different density regions may provide an explanation for the intermittent LPI behaviors observed in the PIC simulations for shock ignition \cite{Yan2014}.

The two-line density profile in Fig. 2(a) were observed in previous experiments with layered targets\cite{Hu2013}. For the parameters in \cite{Hu2013}, the required laser intensity for the gradient-change-induced absolute SRS and TPD are $4\times10^{15}W/cm^2$ and $8\times10^{14}W/cm^2$, respectively. Therefore it is possible to experimentally test the theory here. 

In summary, we found a sufficient change of the density gradient in a linear density profile can turn convective instabilities into absolute ones. An analytical expression is given for the threshold of the density gradient change, which depends on the convective gain only. This theory can explain fluid simulations where a static sinusoidal density modulation is superimposed on a linear density profile and the same convective-to-absolute conversion is also observed. This conversion can occur for TPD and SRS instabilities under realistic direct-drive ICF conditions and can affect hot electron generation and laser energy deposition. The results show that it is important to consider density modulation-induced modes when assessing LPI in ICF. 

This work was supported by DOE under Grant No.
DE-FC02-04ER54789 and DE-SC0012316; DOE NNSA under Award No. DE-NA0001944; by NSF under Grant No.
PHY-1314734; and by National Natural Science Foundation of China
(NSFC) under Grant No. 11129503. The research used resources of the
National Energy Research Scientific Computing Center. The support of
DOE does not constitute an endorsement by DOE of the views expressed
in this paper.


\begin{thebibliography}{25}%
\makeatletter
\providecommand \@ifxundefined [1]{%
 \@ifx{#1\undefined}
}%
\providecommand \@ifnum [1]{%
 \ifnum #1\expandafter \@firstoftwo
 \else \expandafter \@secondoftwo
 \fi
}%
\providecommand \@ifx [1]{%
 \ifx #1\expandafter \@firstoftwo
 \else \expandafter \@secondoftwo
 \fi
}%
\providecommand \natexlab [1]{#1}%
\providecommand \enquote  [1]{``#1''}%
\providecommand \bibnamefont  [1]{#1}%
\providecommand \bibfnamefont [1]{#1}%
\providecommand \citenamefont [1]{#1}%
\providecommand \href@noop [0]{\@secondoftwo}%
\providecommand \href [0]{\begingroup \@sanitize@url \@href}%
\providecommand \@href[1]{\@@startlink{#1}\@@href}%
\providecommand \@@href[1]{\endgroup#1\@@endlink}%
\providecommand \@sanitize@url [0]{\catcode `\\12\catcode `\$12\catcode
  `\&12\catcode `\#12\catcode `\^12\catcode `\_12\catcode `\%12\relax}%
\providecommand \@@startlink[1]{}%
\providecommand \@@endlink[0]{}%
\providecommand \url  [0]{\begingroup\@sanitize@url \@url }%
\providecommand \@url [1]{\endgroup\@href {#1}{\urlprefix }}%
\providecommand \urlprefix  [0]{URL }%
\providecommand \Eprint [0]{\href }%
\providecommand \doibase [0]{http://dx.doi.org/}%
\providecommand \selectlanguage [0]{\@gobble}%
\providecommand \bibinfo  [0]{\@secondoftwo}%
\providecommand \bibfield  [0]{\@secondoftwo}%
\providecommand \translation [1]{[#1]}%
\providecommand \BibitemOpen [0]{}%
\providecommand \bibitemStop [0]{}%
\providecommand \bibitemNoStop [0]{.\EOS\space}%
\providecommand \EOS [0]{\spacefactor3000\relax}%
\providecommand \BibitemShut  [1]{\csname bibitem#1\endcsname}%
\let\auto@bib@innerbib\@empty
%</preamble>
\bibitem [{\citenamefont {Rosenbluth}(1972)}]{Rosenbluth1972}%
  \BibitemOpen
  \bibfield  {author} {\bibinfo {author} {\bibfnamefont {M.~N.}\ \bibnamefont
  {Rosenbluth}},\ }\href@noop {} {\bibfield  {journal} {\bibinfo  {journal}
  {Phys. Rev. Lett.}\ }\textbf {\bibinfo {volume} {29}},\ \bibinfo {pages}
  {565} (\bibinfo {year} {1972})}\BibitemShut {NoStop}%
\bibitem [{\citenamefont {Strozzi}\ \emph {et~al.}(2008)\citenamefont
  {Strozzi}, \citenamefont {Williams}, \citenamefont {Hinkel}, \citenamefont
  {Froula}, \citenamefont {London},\ and\ \citenamefont
  {Callahan}}]{Strozzi2008}%
  \BibitemOpen
  \bibfield  {author} {\bibinfo {author} {\bibfnamefont {D.~J.}\ \bibnamefont
  {Strozzi}}, \bibinfo {author} {\bibfnamefont {E.~A.}\ \bibnamefont
  {Williams}}, \bibinfo {author} {\bibfnamefont {D.~E.}\ \bibnamefont
  {Hinkel}}, \bibinfo {author} {\bibfnamefont {D.~H.}\ \bibnamefont {Froula}},
  \bibinfo {author} {\bibfnamefont {R.~A.}\ \bibnamefont {London}}, \ and\
  \bibinfo {author} {\bibfnamefont {D.~A.}\ \bibnamefont {Callahan}},\ }\href
  {\doibase 10.1063/1.2992522} {\bibfield  {journal} {\bibinfo  {journal}
  {Phys. Plasmas}\ }\textbf {\bibinfo {volume} {15}},\ \bibinfo {pages}
  {102703} (\bibinfo {year} {2008})}\BibitemShut {NoStop}%
\bibitem [{\citenamefont {Nicholson}(1976)}]{Nicholson1976}%
  \BibitemOpen
  \bibfield  {author} {\bibinfo {author} {\bibfnamefont {D.~R.}\ \bibnamefont
  {Nicholson}},\ }\href {\doibase 10.1063/1.861555} {\bibfield  {journal}
  {\bibinfo  {journal} {Phys. Fluids}\ }\textbf {\bibinfo {volume} {19}},\
  \bibinfo {pages} {889} (\bibinfo {year} {1976})}\BibitemShut {NoStop}%
\bibitem [{\citenamefont {Nicholson}\ and\ \citenamefont
  {Kaufman}(1974)}]{Nicholson1974}%
  \BibitemOpen
  \bibfield  {author} {\bibinfo {author} {\bibfnamefont {D.~R.}\ \bibnamefont
  {Nicholson}}\ and\ \bibinfo {author} {\bibfnamefont {A.~N.}\ \bibnamefont
  {Kaufman}},\ }\href@noop {} {\bibfield  {journal} {\bibinfo  {journal} {Phys.
  Rev. Lett.}\ }\textbf {\bibinfo {volume} {33}} (\bibinfo {year}
  {1974})}\BibitemShut {NoStop}%
\bibitem [{\citenamefont {Laval}\ \emph {et~al.}(1976)\citenamefont {Laval},
  \citenamefont {Pellat},\ and\ \citenamefont {Pesme}}]{Laval1976}%
  \BibitemOpen
  \bibfield  {author} {\bibinfo {author} {\bibfnamefont {G.}~\bibnamefont
  {Laval}}, \bibinfo {author} {\bibfnamefont {R.}~\bibnamefont {Pellat}}, \
  and\ \bibinfo {author} {\bibfnamefont {D.}~\bibnamefont {Pesme}},\ }\href
  {\doibase 10.1103/PhysRevLett.36.192} {\bibfield  {journal} {\bibinfo
  {journal} {Phys. Rev. Lett.}\ }\textbf {\bibinfo {volume} {36}},\ \bibinfo
  {pages} {192} (\bibinfo {year} {1976})}\BibitemShut {NoStop}%
\bibitem [{\citenamefont {Picard}\ and\ \citenamefont
  {Johnston}(1983)}]{Picard1983}%
  \BibitemOpen
  \bibfield  {author} {\bibinfo {author} {\bibfnamefont {G.}~\bibnamefont
  {Picard}}\ and\ \bibinfo {author} {\bibfnamefont {T.~W.}\ \bibnamefont
  {Johnston}},\ }\href {\doibase 10.1103/PhysRevLett.51.574} {\bibfield
  {journal} {\bibinfo  {journal} {Phys. Rev. Lett.}\ }\textbf {\bibinfo
  {volume} {51}},\ \bibinfo {pages} {574} (\bibinfo {year} {1983})}\BibitemShut
  {NoStop}%
\bibitem [{\citenamefont {Rosenbluth}\ \emph {et~al.}(1973)\citenamefont
  {Rosenbluth}, \citenamefont {White},\ and\ \citenamefont
  {Liu}}]{Rosenbluth1973}%
  \BibitemOpen
  \bibfield  {author} {\bibinfo {author} {\bibfnamefont {M.~N.}\ \bibnamefont
  {Rosenbluth}}, \bibinfo {author} {\bibfnamefont {R.~B.}\ \bibnamefont
  {White}}, \ and\ \bibinfo {author} {\bibfnamefont {C.~S.}\ \bibnamefont
  {Liu}},\ }\href {\doibase 10.1103/PhysRevLett.31.1190} {\bibfield  {journal}
  {\bibinfo  {journal} {Phys. Rev. Lett.}\ }\textbf {\bibinfo {volume} {31}},\
  \bibinfo {pages} {1190} (\bibinfo {year} {1973})}\BibitemShut {NoStop}%
\bibitem [{\citenamefont {Liu}\ and\ \citenamefont
  {Rosenbluth}(1976)}]{Liu1976}%
  \BibitemOpen
  \bibfield  {author} {\bibinfo {author} {\bibfnamefont {C.~S.}\ \bibnamefont
  {Liu}}\ and\ \bibinfo {author} {\bibfnamefont {M.~N.}\ \bibnamefont
  {Rosenbluth}},\ }\href {\doibase 10.1063/1.861591} {\bibfield  {journal}
  {\bibinfo  {journal} {Phys. Fluids}\ }\textbf {\bibinfo {volume} {19}},\
  \bibinfo {pages} {967} (\bibinfo {year} {1976})}\BibitemShut {NoStop}%
\bibitem [{\citenamefont {Yan}\ \emph {et~al.}(2012)\citenamefont {Yan},
  \citenamefont {Ren}, \citenamefont {Li}, \citenamefont {Maximov},
  \citenamefont {Mori}, \citenamefont {Sheng},\ and\ \citenamefont
  {Tsung}}]{Yan2012}%
  \BibitemOpen
  \bibfield  {author} {\bibinfo {author} {\bibfnamefont {R.}~\bibnamefont
  {Yan}}, \bibinfo {author} {\bibfnamefont {C.}~\bibnamefont {Ren}}, \bibinfo
  {author} {\bibfnamefont {J.}~\bibnamefont {Li}}, \bibinfo {author}
  {\bibfnamefont {A.~V.}\ \bibnamefont {Maximov}}, \bibinfo {author}
  {\bibfnamefont {W.~B.}\ \bibnamefont {Mori}}, \bibinfo {author}
  {\bibfnamefont {Z.~M.}\ \bibnamefont {Sheng}}, \ and\ \bibinfo {author}
  {\bibfnamefont {F.~S.}\ \bibnamefont {Tsung}},\ }\href {\doibase
  10.1103/PhysRevLett.108.175002} {\bibfield  {journal} {\bibinfo  {journal}
  {Phys. Rev. Lett.}\ }\textbf {\bibinfo {volume} {108}},\ \bibinfo {pages}
  {175002} (\bibinfo {year} {2012})}\BibitemShut {NoStop}%
\bibitem [{\citenamefont {Drake}\ \emph {et~al.}(1974)\citenamefont {Drake},
  \citenamefont {Kaw}, \citenamefont {Lee}, \citenamefont {Schmid},
  \citenamefont {Liu},\ and\ \citenamefont {Rosenbluth}}]{Drake1974}%
  \BibitemOpen
  \bibfield  {author} {\bibinfo {author} {\bibfnamefont {J.~F.}\ \bibnamefont
  {Drake}}, \bibinfo {author} {\bibfnamefont {P.~K.}\ \bibnamefont {Kaw}},
  \bibinfo {author} {\bibfnamefont {Y.~C.}\ \bibnamefont {Lee}}, \bibinfo
  {author} {\bibfnamefont {G.}~\bibnamefont {Schmid}}, \bibinfo {author}
  {\bibfnamefont {C.~S.}\ \bibnamefont {Liu}}, \ and\ \bibinfo {author}
  {\bibfnamefont {M.~N.}\ \bibnamefont {Rosenbluth}},\ }\href {\doibase
  http://dx.doi.org/10.1063/1.1694789} {\bibfield  {journal} {\bibinfo
  {journal} {Phys. Fluids}\ }\textbf {\bibinfo {volume} {17}},\ \bibinfo
  {pages} {778} (\bibinfo {year} {1974})}\BibitemShut {NoStop}%
\bibitem [{\citenamefont {Forslund}\ \emph {et~al.}(1975)\citenamefont
  {Forslund}, \citenamefont {Kindel},\ and\ \citenamefont
  {Lindman}}]{Forslund1975}%
  \BibitemOpen
  \bibfield  {author} {\bibinfo {author} {\bibfnamefont {D.~W.}\ \bibnamefont
  {Forslund}}, \bibinfo {author} {\bibfnamefont {J.~M.}\ \bibnamefont
  {Kindel}}, \ and\ \bibinfo {author} {\bibfnamefont {E.~L.}\ \bibnamefont
  {Lindman}},\ }\href {\doibase http://dx.doi.org/10.1063/1.861248} {\bibfield
  {journal} {\bibinfo  {journal} {Phys. Fluids}\ }\textbf {\bibinfo {volume}
  {18}},\ \bibinfo {pages} {1002} (\bibinfo {year} {1975})}\BibitemShut
  {NoStop}%
\bibitem [{\citenamefont {Betti}\ \emph {et~al.}(2007)\citenamefont {Betti},
  \citenamefont {Zhou}, \citenamefont {Anderson}, \citenamefont {Perkins},
  \citenamefont {Theobald},\ and\ \citenamefont {Solodov}}]{Betti2007}%
  \BibitemOpen
  \bibfield  {author} {\bibinfo {author} {\bibfnamefont {R.}~\bibnamefont
  {Betti}}, \bibinfo {author} {\bibfnamefont {C.~D.}\ \bibnamefont {Zhou}},
  \bibinfo {author} {\bibfnamefont {K.~S.}\ \bibnamefont {Anderson}}, \bibinfo
  {author} {\bibfnamefont {L.~J.}\ \bibnamefont {Perkins}}, \bibinfo {author}
  {\bibfnamefont {W.}~\bibnamefont {Theobald}}, \ and\ \bibinfo {author}
  {\bibfnamefont {A.~A.}\ \bibnamefont {Solodov}},\ }\href {\doibase
  10.1103/PhysRevLett.98.155001} {\bibfield  {journal} {\bibinfo  {journal}
  {Phys. Rev. Lett.}\ }\textbf {\bibinfo {volume} {98}},\ \bibinfo {pages}
  {155001} (\bibinfo {year} {2007})}\BibitemShut {NoStop}%
\bibitem [{\citenamefont {Theobald}\ \emph {et~al.}(2012)\citenamefont
  {Theobald}, \citenamefont {Nora}, \citenamefont {Lafon}, \citenamefont
  {Casner}, \citenamefont {Ribeyre}, \citenamefont {Anderson}, \citenamefont
  {Betti}, \citenamefont {Delettrez}, \citenamefont {Frenje}, \citenamefont
  {Glebov}, \citenamefont {Gotchev}, \citenamefont {Hohenberger}, \citenamefont
  {Hu}, \citenamefont {Marshall}, \citenamefont {Meyerhofer}, \citenamefont
  {Sangster}, \citenamefont {Schurtz}, \citenamefont {Seka}, \citenamefont
  {Smalyuk}, \citenamefont {Stoeckl},\ and\ \citenamefont
  {Yaakobi}}]{Theobald2012}%
  \BibitemOpen
  \bibfield  {author} {\bibinfo {author} {\bibfnamefont {W.}~\bibnamefont
  {Theobald}}, \bibinfo {author} {\bibfnamefont {R.}~\bibnamefont {Nora}},
  \bibinfo {author} {\bibfnamefont {M.}~\bibnamefont {Lafon}}, \bibinfo
  {author} {\bibfnamefont {A.}~\bibnamefont {Casner}}, \bibinfo {author}
  {\bibfnamefont {X.}~\bibnamefont {Ribeyre}}, \bibinfo {author} {\bibfnamefont
  {K.~S.}\ \bibnamefont {Anderson}}, \bibinfo {author} {\bibfnamefont
  {R.}~\bibnamefont {Betti}}, \bibinfo {author} {\bibfnamefont {J.~A.}\
  \bibnamefont {Delettrez}}, \bibinfo {author} {\bibfnamefont {J.~A.}\
  \bibnamefont {Frenje}}, \bibinfo {author} {\bibfnamefont {V.~Y.}\
  \bibnamefont {Glebov}}, \bibinfo {author} {\bibfnamefont {O.~V.}\
  \bibnamefont {Gotchev}}, \bibinfo {author} {\bibfnamefont {M.}~\bibnamefont
  {Hohenberger}}, \bibinfo {author} {\bibfnamefont {S.~X.}\ \bibnamefont {Hu}},
  \bibinfo {author} {\bibfnamefont {F.~J.}\ \bibnamefont {Marshall}}, \bibinfo
  {author} {\bibfnamefont {D.~D.}\ \bibnamefont {Meyerhofer}}, \bibinfo
  {author} {\bibfnamefont {T.~C.}\ \bibnamefont {Sangster}}, \bibinfo {author}
  {\bibfnamefont {G.}~\bibnamefont {Schurtz}}, \bibinfo {author} {\bibfnamefont
  {W.}~\bibnamefont {Seka}}, \bibinfo {author} {\bibfnamefont {V.~A.}\
  \bibnamefont {Smalyuk}}, \bibinfo {author} {\bibfnamefont {C.}~\bibnamefont
  {Stoeckl}}, \ and\ \bibinfo {author} {\bibfnamefont {B.}~\bibnamefont
  {Yaakobi}},\ }\href@noop {} {\bibfield  {journal} {\bibinfo  {journal} {Phys.
  Plasmas}\ }\textbf {\bibinfo {volume} {19}} (\bibinfo {year}
  {2012})}\BibitemShut {NoStop}%
\bibitem [{\citenamefont {Yan}\ \emph {et~al.}(2014)\citenamefont {Yan},
  \citenamefont {Li},\ and\ \citenamefont {Ren}}]{Yan2014}%
  \BibitemOpen
  \bibfield  {author} {\bibinfo {author} {\bibfnamefont {R.}~\bibnamefont
  {Yan}}, \bibinfo {author} {\bibfnamefont {J.}~\bibnamefont {Li}}, \ and\
  \bibinfo {author} {\bibfnamefont {C.}~\bibnamefont {Ren}},\ }\href {\doibase
  10.1063/1.4882682} {\bibfield  {journal} {\bibinfo  {journal} {Phys.
  Plasmas}\ }\textbf {\bibinfo {volume} {21}},\ \bibinfo {pages} {062705}
  (\bibinfo {year} {2014})}\BibitemShut {NoStop}%
\bibitem [{\citenamefont {Simon}\ \emph {et~al.}(1983)\citenamefont {Simon},
  \citenamefont {Short}, \citenamefont {Williams},\ and\ \citenamefont
  {Dewandre}}]{Simon1983}%
  \BibitemOpen
  \bibfield  {author} {\bibinfo {author} {\bibfnamefont {A.}~\bibnamefont
  {Simon}}, \bibinfo {author} {\bibfnamefont {R.~W.}\ \bibnamefont {Short}},
  \bibinfo {author} {\bibfnamefont {E.~A.}\ \bibnamefont {Williams}}, \ and\
  \bibinfo {author} {\bibfnamefont {T.}~\bibnamefont {Dewandre}},\ }\href
  {\doibase 10.1063/1.864037} {\bibfield  {journal} {\bibinfo  {journal} {Phys.
  Fluids}\ }\textbf {\bibinfo {volume} {26}},\ \bibinfo {pages} {3107}
  (\bibinfo {year} {1983})}\BibitemShut {NoStop}%
\bibitem [{\citenamefont {Smalyuk}\ \emph {et~al.}(2008)\citenamefont
  {Smalyuk}, \citenamefont {Shvarts}, \citenamefont {Betti}, \citenamefont
  {Delettrez}, \citenamefont {Edgell}, \citenamefont {Glebov}, \citenamefont
  {Goncharov}, \citenamefont {McCrory}, \citenamefont {Meyerhofer},
  \citenamefont {Radha}, \citenamefont {Regan}, \citenamefont {Sangster},
  \citenamefont {Seka}, \citenamefont {Skupsky}, \citenamefont {Stoeckl},
  \citenamefont {Yaakobi}, \citenamefont {Frenje}, \citenamefont {Li},
  \citenamefont {Petrasso},\ and\ \citenamefont {S{\'{e}}guin}}]{Smalyuk2008}%
  \BibitemOpen
  \bibfield  {author} {\bibinfo {author} {\bibfnamefont {V.~A.}\ \bibnamefont
  {Smalyuk}}, \bibinfo {author} {\bibfnamefont {D.}~\bibnamefont {Shvarts}},
  \bibinfo {author} {\bibfnamefont {R.}~\bibnamefont {Betti}}, \bibinfo
  {author} {\bibfnamefont {J.~A.}\ \bibnamefont {Delettrez}}, \bibinfo {author}
  {\bibfnamefont {D.~H.}\ \bibnamefont {Edgell}}, \bibinfo {author}
  {\bibfnamefont {V.~Y.}\ \bibnamefont {Glebov}}, \bibinfo {author}
  {\bibfnamefont {V.~N.}\ \bibnamefont {Goncharov}}, \bibinfo {author}
  {\bibfnamefont {R.~L.}\ \bibnamefont {McCrory}}, \bibinfo {author}
  {\bibfnamefont {D.~D.}\ \bibnamefont {Meyerhofer}}, \bibinfo {author}
  {\bibfnamefont {P.~B.}\ \bibnamefont {Radha}}, \bibinfo {author}
  {\bibfnamefont {S.~P.}\ \bibnamefont {Regan}}, \bibinfo {author}
  {\bibfnamefont {T.~C.}\ \bibnamefont {Sangster}}, \bibinfo {author}
  {\bibfnamefont {W.}~\bibnamefont {Seka}}, \bibinfo {author} {\bibfnamefont
  {S.}~\bibnamefont {Skupsky}}, \bibinfo {author} {\bibfnamefont
  {C.}~\bibnamefont {Stoeckl}}, \bibinfo {author} {\bibfnamefont
  {B.}~\bibnamefont {Yaakobi}}, \bibinfo {author} {\bibfnamefont {J.~A.}\
  \bibnamefont {Frenje}}, \bibinfo {author} {\bibfnamefont {C.~K.}\
  \bibnamefont {Li}}, \bibinfo {author} {\bibfnamefont {R.~D.}\ \bibnamefont
  {Petrasso}}, \ and\ \bibinfo {author} {\bibfnamefont {F.~H.}\ \bibnamefont
  {S{\'{e}}guin}},\ }\href {\doibase 10.1103/PhysRevLett.100.185005} {\bibfield
   {journal} {\bibinfo  {journal} {Phys. Rev. Lett.}\ }\textbf {\bibinfo
  {volume} {100}},\ \bibinfo {pages} {185005} (\bibinfo {year}
  {2008})}\BibitemShut {NoStop}%
\bibitem [{\citenamefont {Goncharov}\ \emph {et~al.}(2008)\citenamefont
  {Goncharov}, \citenamefont {Sangster}, \citenamefont {Radha}, \citenamefont
  {Betti}, \citenamefont {Boehly}, \citenamefont {Collins}, \citenamefont
  {Craxton}, \citenamefont {Delettrez}, \citenamefont {Epstein}, \citenamefont
  {Glebov}, \citenamefont {Hu}, \citenamefont {Igumenshchev}, \citenamefont
  {Knauer}, \citenamefont {Loucks}, \citenamefont {Marozas}, \citenamefont
  {Marshall}, \citenamefont {McCrory}, \citenamefont {McKenty}, \citenamefont
  {Meyerhofer}, \citenamefont {Regan}, \citenamefont {Seka}, \citenamefont
  {Skupsky}, \citenamefont {Smalyuk}, \citenamefont {Soures}, \citenamefont
  {Stoeckl}, \citenamefont {Shvarts}, \citenamefont {Frenje}, \citenamefont
  {Petrasso}, \citenamefont {Li}, \citenamefont {Seguin}, \citenamefont
  {Manheimer},\ and\ \citenamefont {Colombant}}]{Goncharov2008}%
  \BibitemOpen
  \bibfield  {author} {\bibinfo {author} {\bibfnamefont {V.~N.}\ \bibnamefont
  {Goncharov}}, \bibinfo {author} {\bibfnamefont {T.~C.}\ \bibnamefont
  {Sangster}}, \bibinfo {author} {\bibfnamefont {P.~B.}\ \bibnamefont {Radha}},
  \bibinfo {author} {\bibfnamefont {R.}~\bibnamefont {Betti}}, \bibinfo
  {author} {\bibfnamefont {T.~R.}\ \bibnamefont {Boehly}}, \bibinfo {author}
  {\bibfnamefont {T.~J.~B.}\ \bibnamefont {Collins}}, \bibinfo {author}
  {\bibfnamefont {R.~S.}\ \bibnamefont {Craxton}}, \bibinfo {author}
  {\bibfnamefont {J.~A.}\ \bibnamefont {Delettrez}}, \bibinfo {author}
  {\bibfnamefont {R.}~\bibnamefont {Epstein}}, \bibinfo {author} {\bibfnamefont
  {V.~Y.}\ \bibnamefont {Glebov}}, \bibinfo {author} {\bibfnamefont {S.~X.}\
  \bibnamefont {Hu}}, \bibinfo {author} {\bibfnamefont {I.~V.}\ \bibnamefont
  {Igumenshchev}}, \bibinfo {author} {\bibfnamefont {J.~P.}\ \bibnamefont
  {Knauer}}, \bibinfo {author} {\bibfnamefont {S.~J.}\ \bibnamefont {Loucks}},
  \bibinfo {author} {\bibfnamefont {J.~A.}\ \bibnamefont {Marozas}}, \bibinfo
  {author} {\bibfnamefont {F.~J.}\ \bibnamefont {Marshall}}, \bibinfo {author}
  {\bibfnamefont {R.~L.}\ \bibnamefont {McCrory}}, \bibinfo {author}
  {\bibfnamefont {P.~W.}\ \bibnamefont {McKenty}}, \bibinfo {author}
  {\bibfnamefont {D.~D.}\ \bibnamefont {Meyerhofer}}, \bibinfo {author}
  {\bibfnamefont {S.~P.}\ \bibnamefont {Regan}}, \bibinfo {author}
  {\bibfnamefont {W.}~\bibnamefont {Seka}}, \bibinfo {author} {\bibfnamefont
  {S.}~\bibnamefont {Skupsky}}, \bibinfo {author} {\bibfnamefont {V.~A.}\
  \bibnamefont {Smalyuk}}, \bibinfo {author} {\bibfnamefont {J.~M.}\
  \bibnamefont {Soures}}, \bibinfo {author} {\bibfnamefont {C.}~\bibnamefont
  {Stoeckl}}, \bibinfo {author} {\bibfnamefont {D.}~\bibnamefont {Shvarts}},
  \bibinfo {author} {\bibfnamefont {J.~A.}\ \bibnamefont {Frenje}}, \bibinfo
  {author} {\bibfnamefont {R.~D.}\ \bibnamefont {Petrasso}}, \bibinfo {author}
  {\bibfnamefont {C.~K.}\ \bibnamefont {Li}}, \bibinfo {author} {\bibfnamefont
  {F.}~\bibnamefont {Seguin}}, \bibinfo {author} {\bibfnamefont
  {W.}~\bibnamefont {Manheimer}}, \ and\ \bibinfo {author} {\bibfnamefont
  {D.~G.}\ \bibnamefont {Colombant}},\ }\href {\doibase 10.1063/1.2856551}
  {\bibfield  {journal} {\bibinfo  {journal} {Phys. Plasmas}\ }\textbf
  {\bibinfo {volume} {15}},\ \bibinfo {pages} {056310} (\bibinfo {year}
  {2008})}\BibitemShut {NoStop}%
\bibitem [{\citenamefont {Seka}\ \emph {et~al.}(2009)\citenamefont {Seka},
  \citenamefont {Edgell}, \citenamefont {Myatt}, \citenamefont {Maximov},
  \citenamefont {Short}, \citenamefont {Goncharov},\ and\ \citenamefont
  {Baldis}}]{Seka2009}%
  \BibitemOpen
  \bibfield  {author} {\bibinfo {author} {\bibfnamefont {W.}~\bibnamefont
  {Seka}}, \bibinfo {author} {\bibfnamefont {D.~H.}\ \bibnamefont {Edgell}},
  \bibinfo {author} {\bibfnamefont {J.~F.}\ \bibnamefont {Myatt}}, \bibinfo
  {author} {\bibfnamefont {A.~V.}\ \bibnamefont {Maximov}}, \bibinfo {author}
  {\bibfnamefont {R.~W.}\ \bibnamefont {Short}}, \bibinfo {author}
  {\bibfnamefont {V.~N.}\ \bibnamefont {Goncharov}}, \ and\ \bibinfo {author}
  {\bibfnamefont {H.~a.}\ \bibnamefont {Baldis}},\ }\href@noop {} {\bibfield
  {journal} {\bibinfo  {journal} {Phys. Plasmas}\ }\textbf {\bibinfo {volume}
  {16}} (\bibinfo {year} {2009})}\BibitemShut {NoStop}%
\bibitem [{\citenamefont {Yan}\ \emph {et~al.}(2009)\citenamefont {Yan},
  \citenamefont {Maximov}, \citenamefont {Ren},\ and\ \citenamefont
  {Tsung}}]{Yan2009}%
  \BibitemOpen
  \bibfield  {author} {\bibinfo {author} {\bibfnamefont {R.}~\bibnamefont
  {Yan}}, \bibinfo {author} {\bibfnamefont {A.~V.}\ \bibnamefont {Maximov}},
  \bibinfo {author} {\bibfnamefont {C.}~\bibnamefont {Ren}}, \ and\ \bibinfo
  {author} {\bibfnamefont {F.~S.}\ \bibnamefont {Tsung}},\ }\href {\doibase
  10.1103/PhysRevLett.103.175002} {\bibfield  {journal} {\bibinfo  {journal}
  {Phys. Rev. Lett.}\ }\textbf {\bibinfo {volume} {103}},\ \bibinfo {pages}
  {175002} (\bibinfo {year} {2009})}\BibitemShut {NoStop}%
\bibitem [{\citenamefont {Yan}\ \emph {et~al.}(2010)\citenamefont {Yan},
  \citenamefont {Maximov},\ and\ \citenamefont {Ren}}]{Yan2010}%
  \BibitemOpen
  \bibfield  {author} {\bibinfo {author} {\bibfnamefont {R.}~\bibnamefont
  {Yan}}, \bibinfo {author} {\bibfnamefont {A.~V.}\ \bibnamefont {Maximov}}, \
  and\ \bibinfo {author} {\bibfnamefont {C.}~\bibnamefont {Ren}},\ }\href
  {\doibase 10.1063/1.3414350} {\bibfield  {journal} {\bibinfo  {journal}
  {Phys. Plasmas}\ }\textbf {\bibinfo {volume} {17}},\ \bibinfo {pages}
  {052701} (\bibinfo {year} {2010})}\BibitemShut {NoStop}%
\bibitem [{\citenamefont {Afeyan}\ and\ \citenamefont
  {Williams}(1997)}]{Afeyan1997}%
  \BibitemOpen
  \bibfield  {author} {\bibinfo {author} {\bibfnamefont {B.~B.}\ \bibnamefont
  {Afeyan}}\ and\ \bibinfo {author} {\bibfnamefont {E.~a.}\ \bibnamefont
  {Williams}},\ }\href {\doibase 10.1063/1.872506} {\bibfield  {journal}
  {\bibinfo  {journal} {Phys. Plasmas}\ }\textbf {\bibinfo {volume} {4}},\
  \bibinfo {pages} {3803} (\bibinfo {year} {1997})}\BibitemShut {NoStop}%
\bibitem [{\citenamefont {Kruer}(2003)}]{Kruer}%
  \BibitemOpen
  \bibfield  {author} {\bibinfo {author} {\bibfnamefont {W.~L.}\ \bibnamefont
  {Kruer}},\ }\href@noop {} {\emph {\bibinfo {title} {The Physics of Laser
  Plasma Interactions}}}\ (\bibinfo  {publisher} {Westview Press},\ \bibinfo
  {address} {Boulder, CO},\ \bibinfo {year} {2003})\BibitemShut {NoStop}%
\bibitem [{\citenamefont {Short}\ and\ \citenamefont
  {Simon}(2004)}]{Short2004}%
  \BibitemOpen
  \bibfield  {author} {\bibinfo {author} {\bibfnamefont {R.~W.}\ \bibnamefont
  {Short}}\ and\ \bibinfo {author} {\bibfnamefont {A.}~\bibnamefont {Simon}},\
  }\href {\doibase 10.1063/1.1798451} {\bibfield  {journal} {\bibinfo
  {journal} {Phys. Plasmas}\ }\textbf {\bibinfo {volume} {11}},\ \bibinfo
  {pages} {5335} (\bibinfo {year} {2004})}\BibitemShut {NoStop}%
\bibitem [{\citenamefont {Chambers}(1975)}]{ChambersPhD}%
  \BibitemOpen
  \bibfield  {author} {\bibinfo {author} {\bibfnamefont {F.~W.}\ \bibnamefont
  {Chambers}},\ }\href@noop {} {Ph.D. thesis},\ \bibinfo  {school}
  {Massachusetts Institute of Technology} (\bibinfo {year} {1975})\BibitemShut
  {NoStop}%
\bibitem [{\citenamefont {Hu}\ \emph {et~al.}(2013)\citenamefont {Hu},
  \citenamefont {Michel}, \citenamefont {Edgell}, \citenamefont {Froula},
  \citenamefont {Follett}, \citenamefont {Goncharov}, \citenamefont {Myatt},
  \citenamefont {Skupsky},\ and\ \citenamefont {Yaakobi}}]{Hu2013}%
  \BibitemOpen
  \bibfield  {author} {\bibinfo {author} {\bibfnamefont {S.~X.}\ \bibnamefont
  {Hu}}, \bibinfo {author} {\bibfnamefont {D.~T.}\ \bibnamefont {Michel}},
  \bibinfo {author} {\bibfnamefont {D.~H.}\ \bibnamefont {Edgell}}, \bibinfo
  {author} {\bibfnamefont {D.~H.}\ \bibnamefont {Froula}}, \bibinfo {author}
  {\bibfnamefont {R.~K.}\ \bibnamefont {Follett}}, \bibinfo {author}
  {\bibfnamefont {V.~N.}\ \bibnamefont {Goncharov}}, \bibinfo {author}
  {\bibfnamefont {J.~F.}\ \bibnamefont {Myatt}}, \bibinfo {author}
  {\bibfnamefont {S.}~\bibnamefont {Skupsky}}, \ and\ \bibinfo {author}
  {\bibfnamefont {B.}~\bibnamefont {Yaakobi}},\ }\href@noop {} {\bibfield
  {journal} {\bibinfo  {journal} {Phys. Plasmas}\ }\textbf {\bibinfo {volume}
  {20}} (\bibinfo {year} {2013})}\BibitemShut {NoStop}%
\end{thebibliography}
\end{document}